\documentclass[aps,reprint,prl,floatfix,long bibliography]{revtex4-2}
\makeatletter\if@twocolumn\PassOptionsToPackage{switch}{lineno}\else\fi\makeatother

\usepackage{tabulary,graphicx,amsfonts,amsmath,amssymb,amsbsy,dcolumn,bm}
\usepackage[utf8]{inputenc}
\usepackage[T1]{fontenc}
\DeclareUnicodeCharacter{2212}{-}
\usepackage{url,multirow,morefloats,floatflt,cancel,tfrupee}
\makeatletter

\AtBeginDocument{\@ifpackageloaded{textcomp}{}{\usepackage{textcomp}}}
\makeatother
\usepackage{colortbl}
\usepackage{xcolor}
\usepackage{pifont}
\usepackage[nointegrals]{wasysym}
\makeatletter

\def\mcWidth#1{\csname TY@F#1\endcsname+\tabcolsep}

\def\cAlignHack{\rightskip\@flushglue\leftskip\@flushglue\parindent\z@\parfillskip\z@skip}
\def\rAlignHack{\rightskip\z@skip\leftskip\@flushglue \parindent\z@\parfillskip\z@skip}

\@ifundefined{etal}{}{}

\usepackage{ifxetex}
\ifxetex\else\if@twocolumn\@ifpackageloaded{stfloats}{}{\usepackage{dblfloatfix}}\fi\fi

\AtBeginDocument{
\expandafter\ifx\csname eqalign\endcsname\relax
\def\eqalign#1{\null\vcenter{\def\\{\cr}\openup\jot\m@th
  \ialign{\strut$\displaystyle{##}$\hfil&$\displaystyle{{}##}$\hfil
      \crcr#1\crcr}}\,}
\fi
}

\AtBeginDocument{%
  \@ifpackageloaded{endfloat}%
   {\renewcommand\efloat@iwrite[1]{\immediate\expandafter\protected@write\csname efloat@post#1\endcsname{}}}{\newif\ifefloat@tables}%
}%

\def\BreakURLText#1{\@tfor\brk@tempa:=#1\do{\brk@tempa\hskip0pt}}
\let\lt=<
\let\gt=>
\def\processVert{\ifmmode|\else\textbar\fi}

\@ifundefined{subparagraph}{
\def\subparagraph{\@startsection{paragraph}{5}{2\parindent}{0ex plus 0.1ex minus 0.1ex}%
{0ex}{\normalfont\small\itshape}}%
}{}

\newcommand\role[1]{\unskip}
\newcommand\aucollab[1]{\unskip}
  
\@ifundefined{tsGraphicsScaleX}{\gdef\tsGraphicsScaleX{1}}{}
\@ifundefined{tsGraphicsScaleY}{\gdef\tsGraphicsScaleY{.9}}{}
\def\checkGraphicsWidth{\ifdim\Gin@nat@width>\linewidth
	\tsGraphicsScaleX\linewidth\else\Gin@nat@width\fi}

\def\checkGraphicsHeight{\ifdim\Gin@nat@height>.9\textheight
	\tsGraphicsScaleY\textheight\else\Gin@nat@height\fi}

\def\fixFloatSize#1{}
\let\ts@includegraphics\includegraphics

\def\inlinegraphic[#1]#2{{\edef\@tempa{#1}\edef\baseline@shift{\ifx\@tempa\@empty0\else#1\fi}\edef\tempZ{\the\numexpr(\numexpr(\baseline@shift*\f@size/100))}\protect\raisebox{\tempZ pt}{\ts@includegraphics{#2}}}}

\AtBeginDocument{\def\includegraphics{\@ifnextchar[{\ts@includegraphics}{\ts@includegraphics[width=\checkGraphicsWidth,height=\checkGraphicsHeight,keepaspectratio]}}}

\DeclareMathAlphabet{\mathpzc}{OT1}{pzc}{m}{it}

\def\URL#1#2{\@ifundefined{href}{#2}{\href{#1}{#2}}}

\def\UrlOrds{\do\*\do\-\do\~\do\'\do\"\do\-}%
\g@addto@macro{\UrlBreaks}{\UrlOrds}

\edef\fntEncoding{\f@encoding}

\makeatother

\def\style#1#2{#2}

\newif\ifmultipleabstract\multipleabstractfalse%
\newcommand{\texttildeapprox}{{\fontfamily{pcr}\selectfont\texttildelow}}
\begin{document}

\title{Crystal Structure and Phonon Density of States of FeSi up to 120 GPa}

\author{Ravhi S Kumar*}
\affiliation{Department of Physics\unskip, University of Illinois at Chicago\unskip, Chicago\unskip, IL 60607\unskip, USA}

\author{Han Liu}

\author{Quan Li*}
\affiliation{International Center for Computational Method, Key Laboratory of Material Simulation Methods \& Software of Ministry of Education and International Center of Future Science,\unskip, Jilin University\unskip, Changchun 130012\unskip, China}

\author{Yuming Xiao}

\author{Paul Chow}

\author{Yue Meng}
\affiliation{HPCAT, X-ray Science Division\unskip, Argonne National Laboratory\unskip, Lemont\unskip, IL 60439\unskip, USA}

\author{Michael Y. Hu}

\author{Ercan Alp}
\affiliation{Advanced Photon Source\unskip, Argonne National Laboratory\unskip, Argonne\unskip, IL 60439\unskip, USA}

\author{Russell Hemley}
\affiliation{Department of Physics, Chemistry, and Earth \& Environmental Sciences\unskip, University of Illinois Chicago\unskip, Chicago\unskip, IL 60607\unskip, USA}

\author{Changfeng Chen}

\author{Andrew L Cornelius}
\affiliation{Department of Physics and Astronomy\unskip, University of Nevada\unskip, Las Vegas\unskip, NV 89154\unskip, USA}

\author{Zachary Fisk}
\affiliation{Department of Physics and Astronomy\unskip, University of California\unskip, Irvine\unskip, CA 92697\unskip, USA}

\begin{abstract}
The strongly correlated material FeSi exhibits several unusual thermal, magnetic, and structural properties under varying pressure-temperature (P-T) conditions. It is a potential thermoelectric alloy and a material of several geological implications as a possible constituent at the Earth's core-mantle boundary (CMB). The phase transition behavior and lattice dynamics of FeSi under different P-T conditions remain elusive. A previous theoretical work predicted a pressure-induced B20-B2 transition at ambient temperature, yet the transition is only observed at high P-T conditions in the experiments. Furthermore, the closing of the electronic gap due to a dramatic renormalization of the electronic structure and phonon anomalies has been reported based on the density function calculations. In this study, we have performed high pressure powder x-ray diffraction and Nuclear Resonant Inelastic X-ray Scattering (NRIXS) measurements up to 120 GPa to understand the phase stability and the lattice dynamics. Our study shows evidence for a nonhydrostatic stress induced B20-B2 transition in FeSi around 36 GPa for the first time. The Fe partial phonon density of states (PDOS) and thermal parameters were derived from NRIXS measurements up to 120 GPa with the density function theoretical (DFT) calculations. These calculations further predict and are consistent with pressure-induced metallization and band gap closing around 12 GPa.

*Corresponding authors: ravhi@uic.edu,\textbf{\space }liquan777@jlu.edu.cn
\end{abstract}\def\keywordstitle{Keywords}

\maketitle 
    
\section{INTRODUCTION}
Iron monosilicide (FeSi) is a particular example of a strongly correlated electron system with d-electrons, which has attracted much attention because of its unusual anomalous temperature dependencies of magnetic\unskip~\cite{1877241:27768841,1877241:27768900,1877241:27768851}, optical \unskip~\cite{1877241:27768859}, elastic \unskip~\cite{1877241:27768864}, transport properties\unskip~\cite{1877241:27768851,1877241:27768871} and noncentrosymmetric crystal structure. FeSi is a narrow band gap semiconductor \unskip~\cite{1877241:27768853} and a promising material for thermoelectric and solar applications \unskip~\cite{1877241:27768855,1877241:27768812,1877241:27768843}. The Fe\ensuremath{_{3}}Al\ensuremath{_{2}}Si\ensuremath{_{3}} based thermoelectric material shows large conductivity and Seebeck coefficient values in which the FeSi microstructure plays a key role in the observed high-power factor \unskip~\cite{1877241:27768850}. In addition, doping of FeSi with Al is reported to lead to a heavy-Fermion metal with unusual properties \unskip~\cite{1877241:27768877}. Tuning the magnetic properties of a narrow band gap semiconductor and its non-fermi liquid state with impurities have fundamental interest in spintronics \unskip~\cite{1877241:27768817}, nonvolatile memory devices \unskip~\cite{1877241:27768836} and quantum many body physics. FeSi is also used in the fabrication of soft magnetic composite materials \unskip~\cite{1877241:27768894,1877241:27768846} which have importance in the commercial high frequency and high-power electromagnetic applications and motors. Most recent experiments and theoretical observations show topological Fermi arcs, bulk chiral fermions in monosilicides \unskip~\cite{1877241:27768869,1877241:27768903,1877241:27768891,1877241:27768825}. They have further attracted considerable interest in FeSi type isostructural materials and their microstructures. 

The effect of pressure on the B20 to B2 phase transition of FeSi has been a long-standing interest. Furthermore, the lattice vibrations and transformation of the electron spectra in the gap energy region is dominated by the dynamics of iron atoms. Since the d-electrons of iron mainly form the electronic spectrum near the Fermi level \unskip~\cite{1877241:27768862}, the rearrangement of the electronic spectrum due to the closing of the gap and the corresponding release of free-charge carriers affects first the bonding forces of iron atoms and hence reflected in the iron partial phonon density of states (PDOS) \unskip~\cite{1877241:27768819}. Therefore, the study of the correlations between the iron partial phonon spectrum and the electron subsystem under variable external parameters can provide important information about physics of the strong correlations and metal-insulator transition in FeSi. The PDOS is also essential to describing a variety of physical properties that are important in many other contexts, ranging from composition of circumstellar dust \unskip~\cite{1877241:27768890} to electronic transport, and strong electronic correlation effects \unskip~\cite{1877241:27768895}.

Potential applications include understanding its presence in the Earth's core-mantle boundary (CMB) and geoscience perspective \unskip~\cite{1877241:27768831,1877241:27768830,1877241:27768824}. Seismic data reveals a considerable reduction in sound velocities and increase in Poisson's ratio in the ultralow-velocity zone (ULVZ) between outer core and lower mantle \unskip~\cite{1877241:27768872}. Light elements (C, O, S, Si, etc.) and their iron alloys are important constituent materials in the CMB \unskip~\cite{1877241:27768882,1877241:27768880,1877241:27768832,1877241:27768816}; among them iron monosilicide (FeSi) has attracted much interest in the last decade as it was proposed as one of the major light-element iron alloys in CMB \unskip~\cite{1877241:27768858,1877241:27768887,1877241:27768834}. Extensive experimental and theoretical studies have been performed in recent years to understand the properties of FeSi at high pressure and temperature conditions. 

Existing compression data, however, provide mixed results about the structural stability and phase transition. At the ambient condition, FeSi is found to be stable in a B20 type cubic structure (\textit{P2\ensuremath{_{1}}3}). Ab initio calculations suggest that the B20 phase undergoes a pressure-induced transformation at ambient T to a B2 type structure above 13-15 GPa \unskip~\cite{1877241:27768863,1877241:27768826,1877241:27768884,1877241:27768875,1877241:27768849}. This structural transition, however, is not found experimentally as predicted. The B20-B2 transition is only reported above 24 GPa and 1950 K \unskip~\cite{1877241:27768879}. More recently, melting studies have shown a first order phase transition to the B2 structure at P-T conditions as low as 12 GPa and 1700 K \unskip~\cite{1877241:27768820}. The equation of state of FeSi has been studied at RT and HT \unskip~\cite{1877241:27768889,1877241:27768852,1877241:27768829,1877241:27768822,1877241:27768823}, and the pressure-induced phase transition holds considerable significance since the difference in compressibility of the B20 and B2 phases could provide key insights into the role and extent of the contribution by FeSi to the density deficit in the CMB. A more comprehensive study on the high-pressure behavior of FeSi is therefore highly desirable. 

~~~In addition to understanding the electronic correlations, the PDOS further provides insight into the anomalous thermal properties, chemical composition through substantial changes in the Debye temperature, the sound velocity and density relation as has been demonstrated for iron. Few NRIXS experiments on FeSi have been reported to date to megabar pressures \unskip~\cite{1877241:27768814,1877241:27768881,1877241:27768845}. Such measurements will allow a systematic and quantitative description of the key thermodynamic properties of FeSi that can be compared to the established models for understanding strong correlations and in geoscience such as Preliminary Earth Reference Model (PREM). 

~~~Here we report on a combined experimental and theoretical study of the structure and phonon density of states of FeSi at high pressure and high temperature. We have performed a suite of high-pressure x-ray diffraction (XRD) and NRIXS experiments to provide a comprehensive description of FeSi under high P-T conditions. we have also carried out ab initio density functional theory calculations to provide systematic benchmarks and key insights into the mechanism of the structural transformation. Our XRD experiments reveal the pressure induced B20 to B2 phase transition at ambient T, and a strong influence of non-hydrostatic stress that significantly lowers the threshold pressure for the onset transition. The PDOS of FeSi up to 120 GPa at ambient T and the compressional and shear velocities compare well with the PDOS of B2 phase synthesized at high P-T by laser heating, indicating that the sample compressed non-hydrostatically has a predominantly B2 phase at 120 GPa. Furthermore, the relation between compressional-wave velocity and density were also obtained in our experiments. We have further examined the bandgap of FeSi using ab initio methods and show a pressure induced metallization. 
    
\section{EXPERIMENATL AND THEORETICAL METHODS}
~~~The isotopically enriched polycrystalline FeSi sample was prepared under arc melting the constituent starting materials. The well ground starting materials with 95\% isotopically enriched Fe and Si were pressed into small pellets and then loaded into a Cu hearth which was introduced in the arc melter. An ultra-pure Ar atmosphere is maintained in the arc melter to prevent any oxidation while reacting the compositions. The pellets were flipped and heated multiple times to obtain a homogeneous reaction, the resultant ingot was then retrieved and characterized using a Bruker D8 Discover in-house powder x-ray diffraction system. We did not find any secondary phase in the reacted material. Single phased FeSi samples in the B20 phase were then loaded into a 100 \ensuremath{\mu}m hole of a preshaped Be gasket (with no Fe impurities) in three post panoramic diamond anvil cell which could accommodate three avalanche photodiode detectors (APD) detectors with ruby pressure marker for the NRIXS experiments. The NRIXS experiments were performed at the 16 IDD sector of HPCAT of APS, ANL. For the NRIXS experiments, the incident photon energy was tuned to 14.4125 keV, the nuclear resonance energy of \ensuremath{^{57}}Fe. The NRIXS signal was measured using the three APD detectors positioned symmetrically and perpendicular to the direction of the incident x-ray beam close to the sample chamber in the panoramic diamond cell to maximize the counts and signal-to-noise ratio. NRIXS data were collected scanning the incident photon energy, with \ensuremath{\Delta }E = \ensuremath{-}100 to +100 meV from the resonant energy, in steps of 0.5 meV with a monochromator energy resolution of 2.2 meV. 

The HPXRD and laser heating experiments were performed at the high-resolution micro-focused diffraction beam line 16-IDB of HPCAT, APS. We used both symmetric and the panoramic DACs for our experiments. The sequence of the high-pressure diffraction experiments is outlined in detail in the following section. The XRD patterns were collected before and after laser heating \unskip~\cite{1877241:27768897}.

~~~First-principles calculations were carried out using the density functional theory with the Perdew-Burke-Ernzerh generalized gradient approximation (GGA) exchange correlation potential \unskip~\cite{1877241:27768840} as implemented in the VASP code \unskip~\cite{1877241:27768885}. We used the projector augmented wave (PAW) pseudo potential method \unskip~\cite{1877241:27768856} with a plane wave basis set, and we employed a cut-off energy of 400 eV and a 16\ensuremath{\times}16\ensuremath{\times}16 Monkhorst-Pack k-point grid to ensure that enthalpy calculations are converged to better than 1 meV/atom. The phase transformation processes are simulated using the climbing image nudged elastic band (CINEB) method \unskip~\cite{1877241:27768844}, taking 12 intermediate images in addition to the two end-point phases. The phonon density of states was calculated using the direct-force method as implemented in the fropho package \unskip~\cite{1877241:27768892}.

$\begin{array}{l}\overset{}{\underset{}{}}\\\end{array} $

\textbf{III. RESULTS AND DISCUSSION}

$\begin{array}{l}\overset{}{\underset{}{}}\\\end{array} $

\textbf{A.} \textbf{High Pressure XRD Experiments}~

We have performed four separate sets of high-pressure XRD experiments on FeSi under different pressure conditions: (i) hydrostatic compression using He as the pressure transmitting medium at RT, (ii) non-hydrostatic compression (no pressure medium) at RT, and (iii) Laser heating and quenching the sample to obtain the B2 phase and (iv) \textit{In-situ} high P-T laser heating with Ne and Ar as pressure media to determine the phase transition boundary. The XRD patterns of the sample compressed in the He medium showed no appreciable change up to 75 GPa in the P-V plot, indicating the stability of the ambient-pressure B20 phase and the absence of the structural transition predicted by previously reported theoretical calculations \unskip~\cite{1877241:27768863,1877241:27768849,1877241:27768906,1877241:27768813}. Under non-hydrostatic compression, we noticed a shoulder appearing above 36 GPa at around 2-theta angle 12.25{\textdegree}, which corresponds to the (102) peak of the B20 phase in addition to a less intense (001) peak around 8{\textdegree}. A cell parameter of a = 2.661(5) {\fontencoding{T1}\selectfont\'\AA} was obtained for the B2 phase. In the third run, the sample was laser heated to 2300 \ensuremath{\pm} 200 K while under compression. The XRD pattern collected after temperature quench indicates that the onset of B20-B2 transition was observed around 29 (2) GPa, lower than the transition pressure of about 36 GPa observed under non-hydrostatic compression at RT. Experiments were repeated with He and Ne pressure media using in-situ laser heating and the onset of the phase transition was found at the same pressure. The P-V EOS derived from the XRD data collected are shown in Fig. 1 (b). There is a volume collapse of 4.3\% and 5\% in the results of the non-hydrostatic compression and laser heating experiments, respectively. The pressure-volume data is fitted with a standard Birch-Murnaghan equation, and the bulk modulus derived for the B20 phase is 168 (4) GPa, which compares well with earlier reports \unskip~\cite{1877241:27768829}\unskip~\cite{1877241:27768823,1877241:27768901,1877241:27768821,1877241:27768902,1877241:27768857}.
\begin{figure}[h!]
 \centering
  \includegraphics{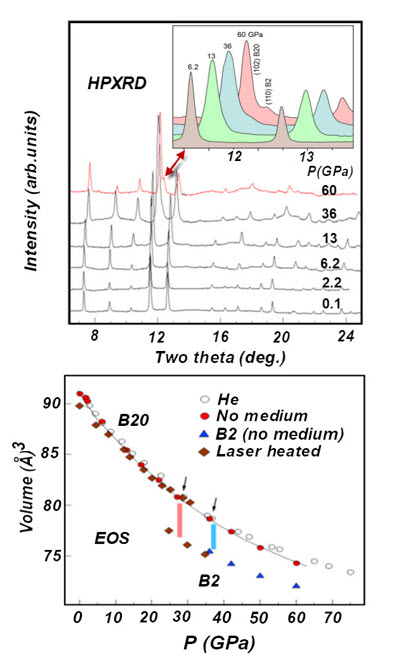}
    \caption{Top: X-ray diffraction patterns collected at various pressures for FeSi under non-hydrostatic compression. The arrow in the inset shows the evolution of the (110) Bragg peak of the B2 phase around 36 GPa and it’s branching out around 60 GPa. Bottom: The pressure-volume data under different compression conditions. Open circles show the P-V data of FeSi with the He medium; solid circles and triangles represent the results under non-hydrostatic compression for the B20 and B2 phase, respectively, obtained from a two-phase fitting of the XRD data; solid diamonds represent the data for the laser heated sample.}
\end{figure}
~~~Our XRD results obtained under different pressure and temperature conditions indicate that the B20-B2 transition is impeded by a high kinetic barrier, which can be overcome by high temperature. This is consistent with previous papers discussing this phenomenon and explains the lack of theoretically predicted phase transition despite the favorable energetics associated with the high-pressure B2 phase of FeSi \unskip~\cite{1877241:27768884,1877241:27768848}. A new result from the present work is the discovery of the crucial role played by non-hydrostatic pressure in triggering the B20-B2 phase transition at much lower pressure compared to hydrostatic compression. The phenomenon of shear-stress induced structural phase transition has received considerable interest in recent years. A plastic shear developed in the DAC could induce a phase transformation that could not be obtained through hydrostatic experiments \unskip~\cite{1877241:27768860} ; several examples are available in the literature such as a superhard wBN transition observed in a disordered nanocrystalline hBN under large plastic shear; first-order rhombohedral-orthorhombic transitions in ferroelectric materials under lattice strain \unskip~\cite{1877241:27768861}, a perovskite to post perovskite phase transition observed in CaIrO\ensuremath{_{3}} under shear stress \unskip~\cite{1877241:27768865} where the layered post-perovskite structure could be directly transformed from the perovskite phase even at room temperature due to high shear stress conditions. Further examples include a e-phase transition of iron induced by shear stress in a rotational diamond anvil cell \unskip~\cite{1877241:27768905} which was not observed under hydrostatic conditions \unskip~\cite{1877241:27768839}. The shear-stress promoted B20-B2 phase transition of FeSi observed in our experiments offers an important case study, and it is consistent with the observation of the strain effects that have been shown to stabilize the B2 phase in FeSi thin films \unskip~\cite{1877241:27768896}. To better understand this phenomenon and explore the underlying mechanism, we performed computational studies of the effect of shear stress as discussed below.

$\begin{array}{l}\overset{}{\underset{}{}}\\\end{array} $

\textbf{B. Effect of Non-hydrostaticity}

$\begin{array}{l}\overset{}{\underset{}{}}\\\end{array} $~The B20 structure has been predicted to become enthalpically unstable with respect to the B2 phase at a moderate pressure of 10-15 GPa \unskip~\cite{1877241:27768884} and 30-40 GPa \unskip~\cite{1877241:27768875}. A plausible cause why this transition was not observed in previous experiments could be due to a large kinetic barrier. Furthermore, previous works \unskip~\cite{1877241:27768907,1877241:27768904}, showed that the structures of FeSi with their competing stability are sensitive to the composition and experimental environment. 
To evaluate the effect of non-hydrostatic stress as a driving force of the kinetic phase-transition process, we have carried out first-principles calculations to evaluate the energetic and kinetic aspects associated with the pressure and strain effects on the B20-B2 phase transition of FeSi. The calculated enthalpy changes along the transformation pathway from the B20 toward the B2 phase at pressures of 20, 60, and 100 GPa are shown in Fig. 2. It is seen that the enthalpy of B2 relative to B20 decreases monotonically with increasing pressure and becomes energetically more favorable when the pressure is higher than 20 GPa. At 60 GPa, the enthalpy of the B2 phase is about 100 meV/atom lower than that of the B20 phase. On the other hand, the kinetic barrier increases with rising pressure, reaching 0.610 eV/atom at 60 GPa, which is much too high to allow phase transitions at ambient temperature. This explains the absence of the energetically favored B2 phase results in many hydrostatic high-pressure experiments. Recently, Niu etal., \unskip~\cite{1877241:27768848} theoretically examined the structural transition under pressure. A comprehensive study of the energetics and kinetics of B20 to B2 phase were studied for a wide pressure range from 4 GPa to 48 GPa. A direct conversion of the B20 phase under hydrostatic compression is predicted to be less likely due to the large conversion barrier between these two phases. They show a possible reconstruction pathway through a metastable Pbcm structure. However, they further point out that the metastable phase exists only for a short pressure range and may not be explicitly observable in the experiments. Besides applying both pressure and temperature that could aid the B20-B2 phase transition, the alternate way is to apply a non-hydrostatic stress. 
\begin{figure}[h!]
 \centering
  \includegraphics{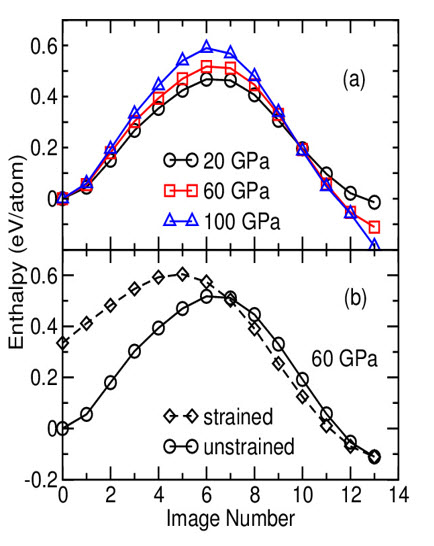}
    \caption{(a) Enthalpy changes relative to the value of B20-FeSi along the pathways for the B20-B2 phase transition at 20, 60, and 100 GPa. (b) Enthalpy change of the unstrained and shear-strained B20-FeSi relative to the value of unstrained B20-FeSi along the pathways for the phase transition toward B2-FeSi at 60 GPa.   To evaluate the kinetic barriers of phase transition between B20 and B2 phases, the phase transformation processes are simulated using 12 intermediate images (structures), in addition to the two end-point structures. Here, image numbers (x) of 0 and 13 correspond to B20 and B2 structures, respectively.}
\end{figure}
To study the influence of the non-hydrostatic pressure condition, we have simulated the kinetic barrier under different stress fields by applying various multi-axial strains in our calculations. We found that shear stress indeed leads to a significant reduction of the kinetic barrier. Fig. 2(b) shows the calculated barriers for the B20-B2 phase transition pathways under unstrained and shear strained pressure conditions at the nominal pressure of 60 GPa when the (110) peak of the B2 phase clearly branches out. To simulate the complex shear strain, the starting B20 phase unit cell is deformed in such a way that the resulting XY, YZ, and ZX stress components are {\texttildeapprox}20 GPa while the XX, YY, and ZZ components are kept at 60 GPa. Our results show that the kinetic barrier is reduced significantly (nearly 33\%) from the original 0.610 eV/atom under the unstrained hydrostatic pressure condition to 0.395 eV/atom under the shear strained. This reduced kinetic barrier is small enough for phase transition to occur at room temperature. This result demonstrates that the B20-B2 transition of FeSi can be induced by the anisotropic stress under non-hydrostatic pressure conditions, which explains the experimental results. Our results add shear stress as another working mechanism, along with the known effect of high temperature that could promote the B20-B2 phase transition.  
$\begin{array}{l}\overset{}{\underset{}{}}\\\end{array} $

\textbf{C. Phonon Density of States ~~~}

The phonon density of states is a crucial component of thermodynamic properties of a material. We performed NRIXS experiments over a wide pressure range from ambient up to 120 GPa and extracted the PDOS by subtracting the elastic peak and fitting the raw NRIXS data as described elsewhere using the PHOENIX software \unskip~\cite{1877241:27768876}. We also simulated the PDOS using first-principles calculations for the B20 and B2 phases of FeSi for comparison with experiment (Fig.3). 
\begin{figure}[h!]
 \centering
  \includegraphics{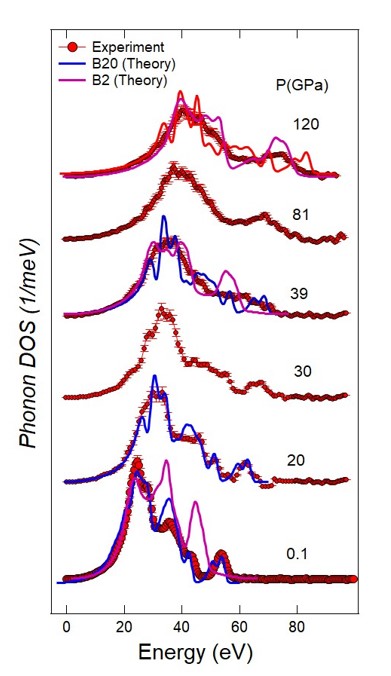}
    \caption{Phonon density of states (DOS) of FeSi at selected pressures.  The solid and dashed lines represent ab initio theoretical results for the B20 and B2 phases, respectively, and the circles with error bars represent the experimental NRIXS data. LH indicates the phonon DOS data collected for the laser heated sample at 17 GPa.}
\end{figure}
The PDOS spectra collected at nearly ambient pressure (0.1 GPa) are well reproduced by the simulated DOS for the B20 phase and are consistent with inelastic neutron, NRIXS measurements \unskip~\cite{1877241:27768819,1877241:27768837}. On increasing pressure, the observed PDOS peaks initially shift systematically to higher energies while the overall spectral features remain essentially the same. The spectra collected at 20 and 30 GPa are well described by the calculated PDOS for the B20 phase with the lattice constant (and volume) contraction being the only changing structural feature under compression. Above 36 GPa, the phonon modes smear out at lower energies and the peaks begin to merge due to further compression of the lattice. At 39 GPa, the experimental data deviates considerably from the calculated results for B2 and B20 phases, indicating the structure is likely in a mixed phase. This is consistent with the results of our XRD measurements that show that the B20 phase starts to transform to the B2 phase above 36 GPa and the PDOS spectra thus likely contain contributions from both the B2 and B20 phases. As pressure is further increased, B2 component is expected to rise and reflected in the experimental PDOS data compared with theoretical calculations. The simulated PDOS at 81 GPa for the B2 phase is found to reproduce the overall phonon spectra observed experimentally. Furthermore, at 120 GPa the simulated PDOS for the B2 phase agrees well with the experimental data. To probe the effect of the kinetic barrier that impedes the B20-to-B2 phase transition, we synthesized a FeSi sample under simultaneous high-pressure and high-temperature (by laser-heating) conditions. Our calculations predict that the B2 phase becomes more stable (lower enthalpy) than the B20 phase at 17 GPa, which is consistent with previous calculations \unskip~\cite{1877241:27768866,1877241:27768867,1877241:27768838,1877241:27768899}. 
\begin{figure}[h!]
 \centering
  \includegraphics{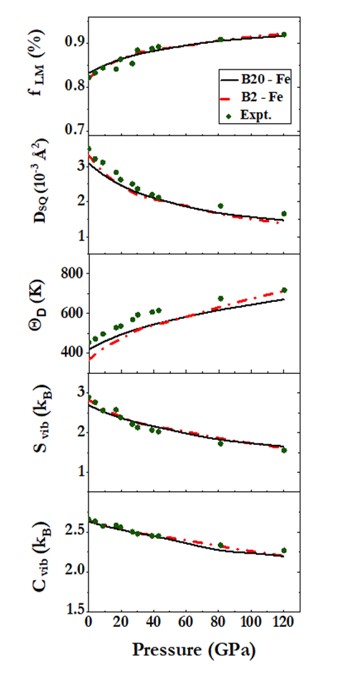}
    \caption{Thermodynamic parameters of FeSi as a function of pressure.  The solid and dashed lines represent the calculated results for the B20 and B2 phase of FeSi, respectively, and the solid circles are the data derived from the measured phonon density of states in our experiments.}
\end{figure}
We further compared (see Fig. 3) the measured PDOS of the FeSi sample synthesized under compression and laser heating which had both B2 and B20 phases. We derive key thermodynamic parameters using the PDOS data, which displays E\ensuremath{^{2 }}scaling behavior. The Debye sound velocity V\ensuremath{_{D}} is obtained from the equation \unskip~\cite{1877241:27768878,1877241:27768883,1877241:27768870,1877241:27768873}

$\begin{array}{l}\begin{array}{l}\overset{}{}D\left(E\right)=(\frac{\widetilde m}m)\frac{E^{2}}{2\pi^{2}nV_D^{3}\ensuremath{\hbar}^{3}}\overset{}{\underset{}{}}\\\end{array}\\\\\end{array} $

Where $\left(\frac{\widetilde m}m\right)\; $= 1.339 is the ratio of the\ensuremath{^{ 57}}Fe atomic mass to the average atomic mass of FeSi and n is the density of atoms. The vibrational heat capacity (C\ensuremath{_{vib}}), entropy (S\ensuremath{_{vib}}), Debye temperature ($\ensuremath{\theta} $\ensuremath{_{D}}), the mean square Displacement (D\ensuremath{_{s}}) and Lamb Mossbauer factor (f\ensuremath{_{LM}}) were derived from the PDOS (Fig. 4). Heat-capacity is calculated via,

$C_{vib}\left(V\right)=k_B{\int{{\left(\frac{\beta E}{2{{sinh}{\left(\frac{\beta E}2\right)}}}\right)}^{2}D\left(E,V\right)dE}} $,
where $\begin{array}{l}\begin{array}{l}\overset{}{\underset{}{}}\\\beta={(k_BT)}^{-1}\overset{}{\underset{}{}}\end{array}\\\\\end{array} $
Entropy can be obtained with the following relation.
\begin{eqnarray*}\style{font-size:12px}{\begin{array}{l}S_{vib}=\frac{k_B\beta}2\int Ecoth\frac{\beta E}2D\left(E,V\right)dE-\overset{}{\underset{}{}}\\k_B\int{{ln(}2sinh\frac{\beta E}2)D\left(E,V\right)dE.}\end{array}} \end{eqnarray*}
Debye temperature is obtained from this equation.

$\begin{array}{l}\overset{}{\underset{}{}}\\\ensuremath{\theta}_D=\frac h{k_B}{\lbrack\frac{3N}{4\pi V}\rbrack}^{1/3}V_D\\\overset{}{\underset{}{}}\\\end{array} $

The mean square Displacement (D\ensuremath{_{s}}) is calculated using the relation \unskip~\cite{1877241:27768827}

$\begin{array}{l}\overset{}{\underset{}{}}\\<D_S>=\frac{E_R}{3k^{2}}\int\frac1Ecoth\frac{\beta E}2D\left(E,V\right)dE\\\overset{}{\underset{}{}}\\\end{array} $.

Lamb Mossbauer factor (f\ensuremath{_{LM}}) is obtained from the equation.

$\begin{array}{l}\overset{}{\underset{}{}}\\f_{LM}=\;e^{-k^{2}<D_S>}\\\overset{}{\underset{}{}}\end{array} $

From the value of V\ensuremath{_{D}}, density \ensuremath{\rho }, and the adiabatic bulk modulus B\ensuremath{_{s}} one can find the aggregate compressional wave velocity V\ensuremath{_{p}}, shear velocity V\ensuremath{_{s}} and shear modulus G from the following equations. 

$\begin{array}{l}\overset{}{\underset{}{}}\\\frac{B_S}\rho=V_P^{2}-\frac43V_S^{2}\\\overset{}{\underset{}{}}\\\end{array} $

$\frac3{V_D^{3}}=\frac1{V_P^{3}}+\frac2{V_S^{3}} $

$\begin{array}{l}\overset{}{\underset{}{}}\\\frac G\rho=\;V_S^{2}\\\overset{}{\underset{}{}}\end{array} $~~~

 The experimental and theoretical data reach a level of agreement that is similar to those previously obtained \unskip~\cite{1877241:27768887,1877241:27768870}. It should be emphasized that the NRIXS measurements here provide the Fe- PDOS, and the theoretical results are compared with thermal parameters. While the partial phonon density of states provides a complete picture in elements, in the case of binary alloys, the contribution from both elements is needed to get the full phonon density of states and thermal parameters. Our calculations indicate significant differences in some of these parameters where the Si contribution plays an important role. We have provided our calculation results for the total FeSi and related details in the supplementary section. Measuring Fe and Si phonon spectra by inelastic neutron scattering and combining the results may provide the complete density of states and corresponding thermal parameters, however, performing neutron scattering experiments and obtaining the phonon spectra at extreme pressures is relatively difficult. 

 Thermal expansion and crystal structure of FeSi investigated over a wide range of temperatures by Vocaldo et al., \unskip~\cite{1877241:27768838} using neutron diffraction found no evidence for structural or magnetic transitions. However, the inelastic neutron scattering experiments by Delaire et al., \unskip~\cite{1877241:27768814} show considerable softening of phonon frequencies with increasing temperature. A crossover from insulating to metallic behavior was reported later due to magnetic fluctuations and strong temperature dependent phonon energies with phonon renormalization leading to a direct spin{\textendash}phonon coupling.~In addition, it has been reported that the acoustic and optical phonon modes of FeSi exhibit different degree of anharmonicity which could be noticed in the non-uniform shifting of various NRIXS modes in the pressure dependency of our experimental phonon spectra. The Debye temperature of 445 (11) K reported earlier agrees well with our experiments and theoretical calculations. We find only a marginal change in the slope of the Debye temperature during the phase transition as the transition is sluggish.
\begin{figure}[h!]
 \centering
  \includegraphics{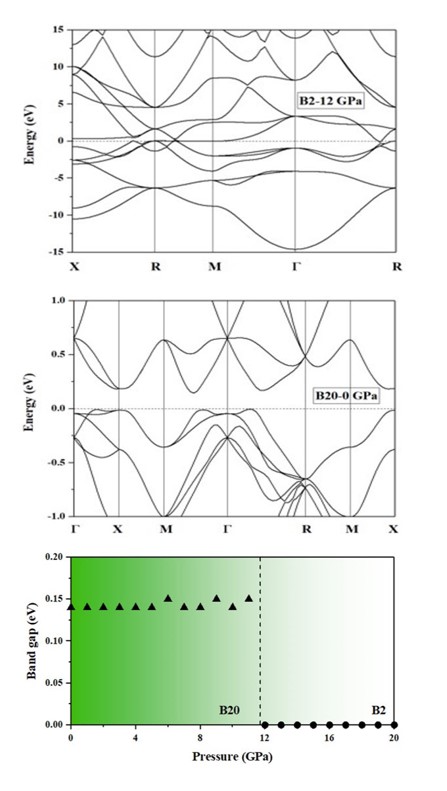}
    \caption{Calculated electronic band structures for FeSi and the band gap as a function of pressure.}
\end{figure}
The calculated band gap and electronic structure as a function of pressure is shown in Fig.5. Above 12 GPa, we could observe a pressure induced metallization in FeSi \unskip~\cite{1877241:27768835,1877241:27768815}. Earlier high pressure studies on FeSi and Ge doped FeSi indicate that the localized states in the gap delocalize, leading to a possible continuous insulator to metal transition below 10 GPa \unskip~\cite{1877241:27768847}. Recent transport studies under compression indicate that the transition to a metallic phase is an electronic change and sensitive to non-hydrostatic compression. The B20-type structure is retained up to 30 GPa, with no evidence of a discontinuity in the volume-pressure equation of state data \unskip~\cite{1877241:27768893} around metallization as observed in our theoretical simulations and x-ray diffraction studies.\textbf{\space }

$\begin{array}{l}\overset{}{\underset{}{}}\\\end{array} $

\textbf{D. Geophysical Implications}

We now discuss the behavior of sound velocities for FeSi based on PDOS and their geological implications. Fe-Si alloys could form as a reaction product between the MgSiO\ensuremath{_{3}} perovskites of the lower mantle with iron in the outer core. Their crystal structure and phase transition at high pressure vary widely depending on the Si content and the P-T conditions. Even though the Si content relevant to the Earth's core that satisfies the seismological constraints is below 10\%, the phase relations of Si-rich Fe-alloys significantly influence the formation of CMB. Silicon content as high as 20\% in weight has been predicted \unskip~\cite{1877241:27768868}. The Si-rich Fe alloys show a stable B2 phase and Si-poor Fe alloys show hcp phase at high pressure. Lin et al. showed that the hcp phase of 8\% Si alloy (Fe\ensuremath{_{0.85}}Si\ensuremath{_{0.15}}) becomes stable above 170 GPa and 3,000 K \unskip~\cite{1877241:27768898}. The Fe-Si alloy with 16\% Si shows a mixture of bcc and hcp phases at high pressure \unskip~\cite{1877241:27768874}. Based on theoretical simulations, higher Si percentages in Fe-Si alloys should stabilize the bcc (B2) phase at high pressure \unskip~\cite{1877241:27768888}. Stoichiometric FeSi (33\% Si) falls in the Si-rich region and the B2 transition is theoretically predicted to occur at pressures below 30 GPa \unskip~\cite{1877241:27768884,1877241:27768875}. 
\begin{figure}[h!]
 \centering
  \includegraphics{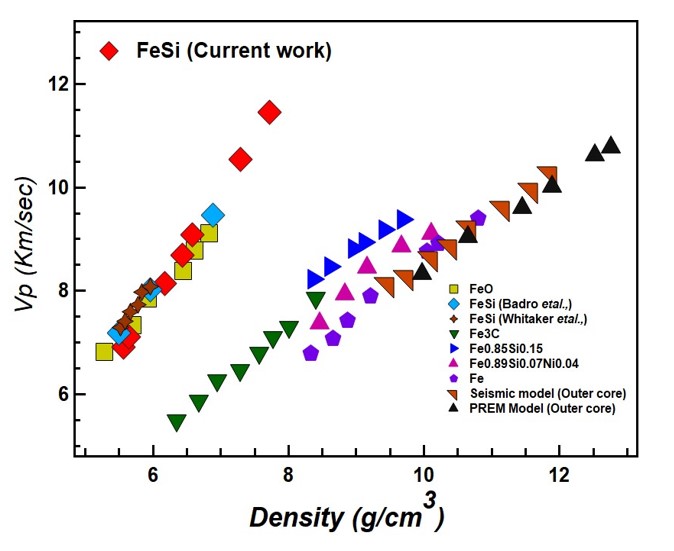}
    \caption{Compressional wave velocity (V\ensuremath{_{p}}) versus density relation for FeSi compared with the results for other Fe alloys.}
\end{figure}
In Fig. 6 we plot the compressional and shear wave velocities of FeSi as a function of pressure. The variation of V\ensuremath{_{p}} as a function of density and pressure are crucial to understanding the materials relevant to the planetary cores. The V\ensuremath{_{p}} data and outer core density profiles obtained by PREM, and the seismic models suggest that the abundance of light elements such as silicon, sulphur and oxygen alloyed with iron at the outer core are constrained by the bulk modulus and the density. Sulphur and nickel cannot be the only major elements alloying with iron, as they would require very high concentrations to account for the density profiles predicted by PREM. In comparison, with silicon, the slope of V\ensuremath{_{p}} is less affected by nickel and sulfur concentration. The data for FeO falls parallel to FeSi; where the partition coefficient measurements show depleted oxygen as an alloying element \unskip~\cite{1877241:27768828}. Based on these imposed constraints, V\ensuremath{_{p}}, and the density data, silicon is proposed to be a viable candidate with 2.8 wt \% at the outer core and 2.3 wt \% in the inner core with trace amounts of oxygen \unskip~\cite{1877241:27768842}. The solubility studies of silicon with iron and perovskites by Takefuji et al., \unskip~\cite{1877241:27768854}\ensuremath{^{}} and the preferred core model proposed by Badro et al., \unskip~\cite{1877241:27768886} agree well with the theoretical calculations on chemical equilibrium of geochemical core conditions and the deviation of the slope of the V\ensuremath{_{p}} vs density data for stoichiometric FeSi with the change of V\ensuremath{_{p}} with increasing atomic weight percentage of silicon. The changes of both V\ensuremath{_{p}} and V\ensuremath{_{s}} as a function of pressure follow a similar trend as those required by PREM up to 120 GPa near the CMB. The density of FeSi is found to be higher than that of PREM at upper mantle, and around CMB, the density matches with the value of PREM as it increases in the ULVZ. The ratio of V\ensuremath{_{p}}\ensuremath{_{}} to V\ensuremath{_{s}}\ensuremath{_{}} of 2.2 calculated from our data at 120 GPa matches well the value expected for ULVZ \unskip~\cite{1877241:27768818,1877241:27768833}.
 $\begin{array}{l}\overset{}{\underset{}{}}\\\end{array} $
\section*{conclusions}
 
 We have studied the structural phase transition of FeSi up to 75 GPa and the phonon density of states up to 120 GPa under hydrostatic and non-hydrostatic compression. We show for the first time that under non-hydrostatic conditions the B20-B2 phase transition is triggered around 36 GPa at room temperature. Theoretical simulations reveal significant shear-stress induced reduction of kinetic barrier that promotes the phase transition of FeSi. Under laser heating, the B20-B2 transition occurs at a much lower pressure of 29 GPa which confirms the theoretical prediction of the relative energetics these FeSi phases and highlight the role of a kinetic barrier in impeding the transition at room temperature as seen in previous and our current experiments. These results explain and reconcile the scattered compression data on FeSi and expand the understanding of the structural transformation of FeSi under diverse compression conditions. We have presented the simulation results showing the band gap closing for FeSi around 12 GPa which agrees with the previously reported results. The phonon density of states obtained from our NRIXS experiments and ab initio calculations further allow a systematic assessment of the compressional wave velocity versus density relationship.  
\section*{Acknowledgments}We thank Tkachev (GSE-CARS) for sample gas loading. Portions of this work were performed at HPCAT (Sector 16), Advanced Photon Source (APS) at Argonne National Laboratory. HPCAT operation is supported by DOE-NNSA's Office of Experimental Sciences under Award DE-NA0001974, DOE-BES under Award DE-FG02-99ER45775, and by partial instrumentation funding from NSF. APS is supported by DOE-BES under Contract DE-AC02-06CH11357. This research used resources of the Advanced Photon Source, a U.S. Department of Energy (DOE) Office of Science user facility and is based on work supported by Laboratory Directed Research and Development (LDRD) funding from Argonne National Laboratory, provided by the Director, Office of Science, of the U.S. DOE under Contract No. DE-AC02-06CH11357. CDAC is supported by DOE-NNSA (DE-NA0003975). 
\appendix
\hfill
\renewcommand{\thefigure}{S\arabic{figure}}
\section{Supplementary Material}
Theoretical calculations show the contributions from Fe and Si to the total phonon DOS for the B2 phase of FeSi (Fig.S1). While analyzing the individual contributions, the Fe weighted part is higher below 35 meV and the Si weighted contributions become dominant above 35 meV. 
\setcounter{figure}{0}    
\begin{table}[h!]
 \centering
  \includegraphics{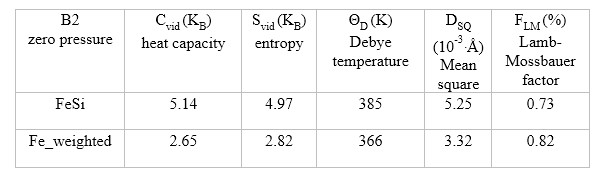}
    \caption{Comparison of the thermodynamic properties from theoretical total (full) phonon DOS and theoretical Fe-weighted phonon DOS for FeSi-B2 phase at zero pressure.}
\end{table}
\begin{figure}[h!]
 \centering
  \includegraphics{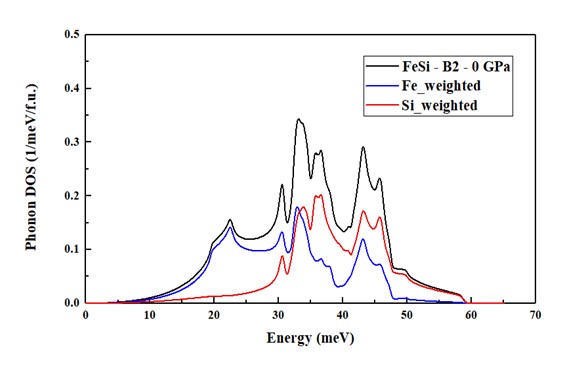}
    \caption{The calculated full phonon DOS, the Fe-weighted phonon DOS, and the Si-weighted phonon DOS for FeSi-B2 phase at zero pressure.}
\end{figure}

The thermodynamic parameters were calculated using a direct integration approach to obtain the phonon DOS, Lamb-Mossbauer factor, the mean square displacement, Debye temperature, entropy, and heat-capacity [1]. As noticed in Table.1., the thermal parameters vary significantly based on the Fe and Si contributions. 
\begin{figure}[h!]
 \centering
  \includegraphics{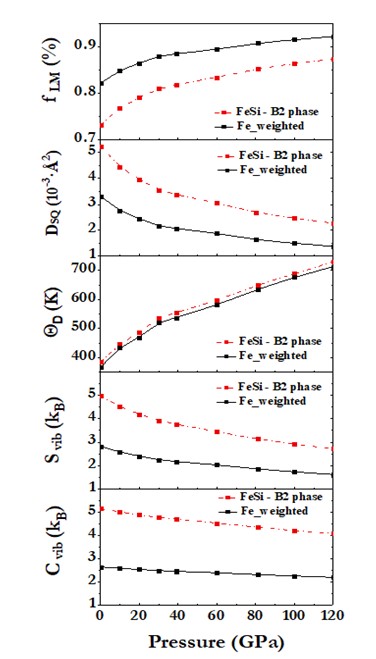}
    \caption{Thermodynamic parameters calculated at various pressures for the B2 phase of FeSi.}
\end{figure}
The thermodynamic parameters are shown for selected pressure in Fig.S2. The Debye temperature showed marginal difference in the values. In NRIXS experiments, the measured quantity is based on the one-phonon contribution which is used to determine the phonon DOS and the Debye sound velocity is obtained from fits to the low energy region of the phonon DOS and the density of the material. However, it could be noticed that all other parameters are also affected due to different contributions from Fe and Si. Consideration of the sensitivity of these parameters by the PDOS fitting and the constituent elements becomes more important in describing these quantities for binary alloys.
[1]. Sturhahn,W., Jackson,J. M., and Ohtani, E. in Advances in High-Pressure Mineralogy Vol. 421 (Geological Society of America, 2007).
\end{document}